\documentclass[
    ,final            
  ]
  {aipproc}

\layoutstyle{8x11double}

\begin{document}

\title{Did supermassive black holes form by direct collapse?}

\classification{<Replace this text with PACS numbers; choose from this list:
                \texttt{http://www.aip..org/pacs/index.html}>}
\keywords      {black hole physics --- accretion, accretion discs --- galaxies: nuclei --- quasars: general}
 
\author{Mitchell C. Begelman}{
  address={JILA, 440 UCB, University of Colorado at Boulder, Boulder, CO 80309-0440 USA}
}

\begin{abstract}
Rapid infall of gas in the nuclei of galaxies could lead to the formation of black holes by direct collapse, without first forming stars.  Black holes formed in this way would have initial masses of a few $M_\odot$, but would be embedded in massive envelopes that would allow them to grow at a highly super-Eddington rate.  Thus, seed black holes as large as $10^3-10^4 M_\odot$ could form very rapidly.  I will sketch the basic physics of the direct collapse process and the properties of the accreting envelopes.
\end{abstract}

\maketitle


\section{Introduction}

Despite years of study, we still do not know how the seeds of supermassive black holes formed.  Few if any of the pathways in Martin Rees's famous flow chart (Begelman \& Rees 1978) can be ruled out, but none of the routes is particularly well understood, either.  What we do know is that some very massive ($> 10^9 M_\odot$) black holes had to exist by $z\sim 6$ in order to explain early quasars (Fan 2006).  If the seeds of these black holes were the remnants of massive stars, then they must have grown by Eddington-limited accretion for most of the time since their formation, or else much of their growth was due to mergers.  A second possibility is that the seeds formed by such a rapid accumulation of matter that it may be considered to be a direct collapse.  I will focus on the latter possibility in this paper.

\section{Direct collapse}

The Pop III star formation processes we have heard about at this conference result from the infall of gas at rates $\sim 10^{-4} - 10^{-2} M_\odot$ yr$^{-1}$.  What would happen if the infall rate were much higher?  The entropy of matter laid down by gravitational infall onto a growing central mass increases with time.  At low inflow rates, however, nuclear ignition halts the contraction of the core and raises the entropy in the interior, leading to a high-entropy object --- a star. If the inflow rate is high enough, however, the core will be so tightly bound by the time nuclear reactions start that the energy release will be insufficient to halt core contraction. In this case, we are left with an object with a low-entropy core and a high-entropy envelope.  This is the situation that can lead to the direct formation of a black hole, without a stellar precursor. This situation should apply when the infall rate exceeds a few tenths of a solar mass per year (Begelman et al.~2006), although more work needs to be done to refine this estimate.  

\subsection{Inflow rate}

The conditions under which such high inflow rates might occur are very uncertain.  Such rapid infall would almost certainly be driven by gravitational torques, which could be local (Gammie 2001) or global (as in the ``bars within bars" mechanism: Shlosman et al.~1989).  The ``natural" gravitational inflow rate is given by $\dot M \sim v^3/G \sim 0.1 (v/10 \ {\rm km \ s}^{-1})^3 M_\odot$ yr$^{-1}$, where $v$ represents the internal velocity dispersion ($\sim$ turbulent or sound speed) for a locally unstable thin disk and the orbital speed for a globally unstable (fully self-gravitating) system. This means that very large inflow rates are possible in dark matter haloes with velocity dispersions exceeding about 10 km s$^{-1}$, which have masses exceeding $\sim 10^8-10^9 M_\odot$ and become common at redshifts $\sim 10-15$.  Global gravitational instabilities could occur in a significant fraction of such haloes if the gas is unable to cool much below the virial temperature ($\sim 10^4$ K), which requires both that they have been spared significant metal enrichment and that H$_2$ formation is suppressed (Bromm \& Loeb 2003; Begelman et al.~2006).  However, recent calculations by Wise \& Abel (2007, and these proceedings) suggest that molecular hydrogen formation is inevitable. This may have the additional effect that many more haloes formed Pop III stars at higher redshifts, making it harder to avoid metal enrichment and possibly depleting the supply of gas available for infall as haloes merge. 

Rapid infall may be easier in more massive haloes. There will be a larger disparity between the thermal temperature of the gas and the virial temperature, but presumably the gas will form a multiphase structure with turbulent velocities (perhaps driven by stellar energy sources) dominating the internal energy --- much like the interstellar medium.  Simulations are beginning to elucidate these structures, but we have a long way to go before we understand the tradeoffs between large scale inflow and in situ fragmentation and star formation.  Inflow may be stimulated by the large-scale gravitational torques associated with mergers. a key element in the simulations of black-hole growth and fuelling (e.g., Di Matteo et al.~2005).  Finally, we note that such inflow rates must be possible, because they are required to power quasars.  The open question is whether they can also occur when the black hole is absent, or very small.  For the remainder of this article we will assume that the answer is affirmative.   

\subsection{Black hole formation}

Once the mass of accumulated gas exceeds a few solar masses, radiation pressure dominates the envelope.  The core, with gas pressure comparable to radiation pressure, maintains a roughly constant mass $\sim 10 M_\odot$.  The boundary between dynamic infall and quasistatic contraction is close to the radius where infalling gas liberates energy at the Eddington limit.  For a constant infall rate of $\dot M = 0.1 
\dot m_{-1} M_\odot  {\rm yr}^{-1}$, this radius turns out to be constant, $R_{\rm env} \sim 1 \dot m_{-1}$ AU.  The core shrinks under the increasing pressure of the envelope, $R_{\rm core} \sim R_{\rm env}/ m_{\rm env}$, where $m_{\rm env} \sim 0.1 \dot m_{-1} t_{\rm yr}$ is the envelope mass in solar units.  

One can sketch the likely interior structure of the envelope using scaling arguments.  Since the interior is undergoing Kelvin--Helmholtz contraction, the density profile adjusts so that the diffusion timescale $t_{\rm diff} \sim (\rho \kappa r^2 /c ) (T/r|\nabla T|)$ is of order the elapsed time at all radii.  This means that the specific entropy of any mass shell, $s(M)$, declines slowly (logarithmically) with time, and is never far below the value it had when the mass shell was added. In the radiation pressure-dominated envelope, $s(M) \propto p_r/p_g \propto M(r)^{1/2}$, where $p_r \propto T^4$ and $p_g \propto \rho T$ are the radiation and gas pressure, respectively.  There are then two possible scaling laws for the envelope structure.  If $T/r | \nabla T| \sim O(1)$, then $\rho(r) \propto r^{-2}$, $T(r) \propto r^{-1/2}$ (for uniform opacity $\kappa$), and the envelope joins smoothly onto the core. This was the result presented in Begelman et al.~(2006). The other possible scaling law 
has $ T/r |\nabla T| \gg 1$, i.e., nearly isothermal structure.  The density profile in the envelope is then $\rho(r) \propto r^{-1/2}$, and there is a large jump in density and temperature going from the envelope to the core.  The latter structure seems to be the one that emerges naturally from ``nearly self-similar" models for the core+envelope, which neglect rotation.  Given the manner in which the mass accumulates, though, rotation is bound to be important. Moreover, even a small amount of it can affect the structure dramatically, increasing the binding energy of the gas (which is otherwise very weakly bound since the mean equation of state has $\gamma\approx 4/3$) and protecting it from pulsational instabilities.  Including the effects of rotation will require the modeling of angular momentum transport and dissipation, but these complications are not prohibitive.  It will be interesting to see which of the scalings applies when rotation is included. 

Fortunately, the behavior of the core is insensitive to the structure of the envelope.  Since the core is largely dominated by gas pressure, its temperature must track the virial temperature, $T_{\rm core} \propto m_{\rm env}$. Rapid nuclear burning starts in the core when the envelope mass is $\sim 100 M_\odot$ and the density is not that different from the cores of main sequence stars, but the system passes through this phase so quickly ($t_{\rm evol} \sim 10^3 \dot m_{-1}^{-1}$ yr) that the energy release has little effect.  By the time the envelope mass reaches several thousand solar masses, the core temperature has climbed to a few $\times 10^8$ K.  Because of the extremely deep potential well created by the envelope, the energy released by nuclear reactions at this stage is unable to unbind the gas (especially if rotation has increased its binding energy) and the core proceeds to the temperatures ($\sim 5 \times 10^8$ K) at which runaway neutrino losses occur.  At this point the core loses pressure support and collapses to form a $\sim 10-20 M_\odot$ black hole.     

\section{Post-BH evolution}

\subsection{Super-Eddington accretion}

At the time of its formation, the black hole is embedded in an envelope of more than a hundred times its mass.  If it were limited to accreting at the Eddington limit for the mass of the black hole, the pressure of the escaping radiation would have essentially no effect on the envelope.  In fact, the pressure of the envelope should be able to drive the accretion rate up to the point where the liberated luminosity approaches the Eddington limit {\it for the mass of the envelope}.  Relative to the mass of the black hole, the accretion is super-Eddington by a factor $ M_{\rm env}/M_{\rm BH}$.

Accretion inside the massive envelope (which itself continues to grow at a rate $\dot M$) can lead to very rapid growth of the black hole. The Eddington ratio for black hole growth is 
\begin{equation}
{\dot M_{\rm BH} \over \dot M_E} \sim 3\times 10^3 \left( {\dot m_{-1}\over \varepsilon_{-1}} \right)^{1/2}  \left( { M_{\rm BH}\over M_\odot} \right)^{-1/2} , 
\end{equation}       
where $0.1\varepsilon_{-1}$ is  the accretion efficiency.  For an initial black hole mass of $10M_\odot$, the e-folding time is a thousandth of the Salpeter time, or only $\sim 10^4$ yr, and super-Eddington growth appears possible up to black hole masses $> 10^6 M_\odot$.  However, this estimate does not take fully into account the back reaction of black hole growth on the envelope.  We will see below that the era of rapid growth is limited to much smaller (but still interesting!) black hole masses. 
 
\subsection{Quasistars}

The energy liberated by accretion has to escape.  Since the accretion flow is rotating, some of it could exhaust through a low-density funnel. But much of it  presumably percolates through the accretion flow and envelope.  Thus, the accreting black hole provides an energy source for the envelope, which is therefore a kind of star-like object which we have dubbed a ``quasistar" (Begelman et al.~2006).  Since there is a limit to the outward energy flux that the accreting gas can carry, the accretion rate is regulated at a fraction $\sim \varepsilon^{-1} (c_s/ c)^2$ of the Bondi rate, where $c_s \sim (p_r/\rho)^{1/2}$ is the sound speed at the Bondi radius (Gruzinov 1998; Blandford \& Begelman 1999; Narayan et al.~2000; Quataert \& Gruzinov 2000).  The accretion rate, and associated energy flux, can be expressed in terms of the temperature and density deep within the quasistar, but outside the black hole's sphere of influence.  The energy flux must also equal the Eddington limit for the entire quasistar, $L_E(M_*)$ (where we now use $M_*$ to denote the mass of the quasistar) --- the interior structure adjusts so that this is satisfied.  Since the Eddington limit is lower inside the quasistar (where the enclosed mass is lower), the quasistar's interior is strongly convective and can be modeled by an $n=3$ polytrope.  (Strictly speaking, one should use a ``loaded polytrope" model [Huntley \& Saslaw 1975], taking into account the mass of the black hole, but the loaded and unloaded models converge in the region of interest when $M_* \gg M_{\rm BH}$.) The interior structure of a quasistar is shown schematically in Fig.~1.

\begin{figure}
  \includegraphics[height=.1\textheight]{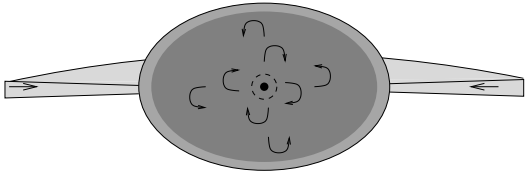}
  \caption{Schematic illustration of the quasistar, from Begelman et al.~(2007). A seed black hole of mass $M_{BH}$ accretes gas from a massive, radiation pressure-supported envelope at a rate set by conditions outside the Bondi radius. The luminosity liberated by the accretion process is transported convectively in the inner regions of the envelope, with a transition to a radiative zone where convection becomes inefficient.  We illustrate rotational flattening and ongoing disk accretion at a fraction of a Solar mass per year.}
\end{figure}

According to the polytropic relations, the density and temperature outside the Bondi radius are uniquely related to the mass and radius ($R_*$) of the quasistar.  Expressing the accretion rate in terms of these quantities and setting the luminosity equal to $L_E(M_*)$, we can express $R_*$ in terms of $M_*$ and $M_{\rm BH}$:
\begin{equation} 
R_* \sim 4 \times 10^{14}  \alpha^{2/5} m_{\rm BH}^{4/5} m_*^{-1/5}  \ {\rm cm}  ,
\end{equation} 
where the masses are expressed in solar units and $\alpha <1$ parametrizes the efficiency of angular momentum and energy transport inside the Bondi radius (Begelman et al.~2007).  The photospheric temperature is
\begin{equation} 
T_{\rm ph} \sim 10^3 \alpha^{-1/5} m_{\rm BH}^{-2/5} m_*^{7/20}  \ {\rm K} .  
\end{equation}

Since the black hole growth rate is proportional to $M_*$, while the quasistar mass increases linearly with time, $M_{\rm BH} \propto M_*^2$ at late times.  This implies that the quasistar's radius grows with time, and its photosphere becomes cooler.  (The result holds even if the quasistar mass is fixed.) The appearance of a quasistar differs dramatically from that of the pre-black hole envelope.  Shortly after the black hole forms the envelope expands to $\sim 100$ AU (from $\sim 1$ AU), and the interior temperature drops to $\sim 10^6$ K, quenching all nuclear reactions. A quasistar resembles a red supergiant, except that it is radiation pressure supported and its energy source is accretion.  In some respects it is reminiscent of a very massive Thorne-\.Zytkow (1977) object, but with crucial differences. Besides being dominated by radiation pressure and powered exclusively by accretion, quasistars have a distributed rather than shell-like energy source, which cannot be regulated by the slow settling that characterizes accretion onto a neutron star.  As a result, quasistars come to grief if their photospheric temperatures get too low.

\begin{figure}
  \includegraphics[height=.5\textheight]{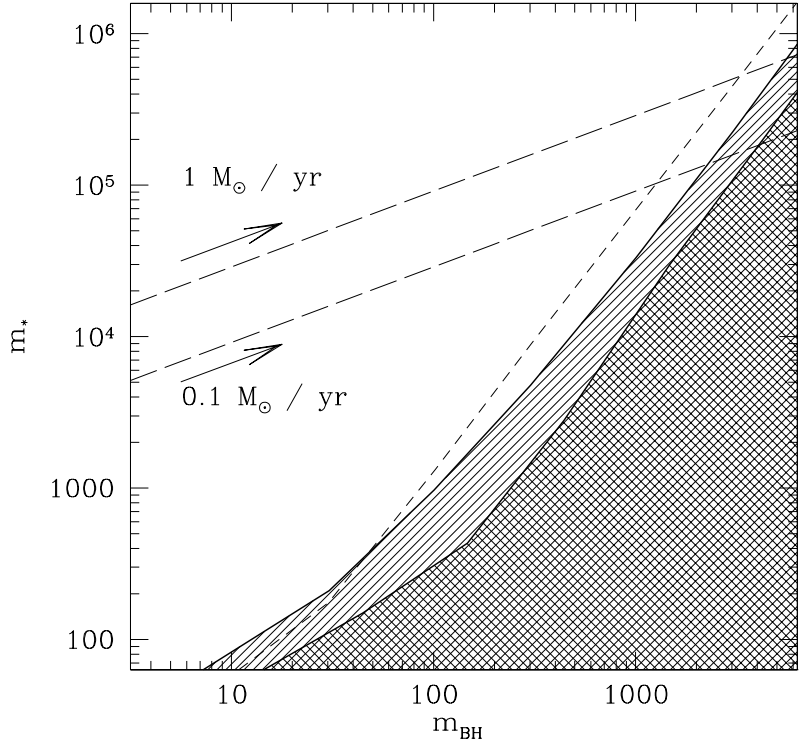}
  \caption{The shaded areas show forbidden photospheric temperatures of Pop III quasistars, as a function of envelope and black hole mass, from Begelman et al.~(2007). The lighter shaded region is computed assuming that $\alpha =0.1$, the darker shaded region is for $\alpha =0.05$, while the dashed line is for an analytic ``toy" opacity model.  Superimposed on the figure are evolutionary tracks for an accretion rate onto the envelope of $\dot{M}_* $= 1 $\ M_{\odot} \ {\rm yr}^{-1}$ and $\dot{M}_*$= 0.1$ \ M_{\odot}\ {\rm yr}^{-1}$.  The evolution pushes the quasistar into the forbidden region of parameter space, where it evaporates.}
\end{figure}

Begelman et al.~(2007) computed quasistar models using Pop III opacities from Mayer \& Duschl (2005).  As in red giants with standard abundances, there is a minimum photospheric temperature associated with a sharp drop in opacity. Because of the absence of metals, the ``opacity crisis" occurs at a somewhat higher temperature --- around 4000 K --- than for metal-enriched compositions.  Like the Hayashi track (Hayashi 1961; Hayashi \& Hoshi 1961) for red giants and convective protostars, the minimum temperature of quasistars arises from the impossibility of matching the convective interior to the radiative zone and photosphere.  The details are somewhat different, however, because radiation pressure dominates in quasistars, whereas the convective zones of ordinary red giants are gas pressure-dominated and resemble $n=3/2$ polytropes.  As a quasistar crosses into the forbidden zone, the flux escaping from the convective interior exceeds the Eddington limit and prevents the quasistar from maintaining hydrostatic equilibrium. Fig.~2 shows the forbidden zone in the $M_*-M_{\rm BH}$ plane, along with representative evolutionary tracks.  

Unlike ordinary red giants reaching the Hayashi track, quasistar photospheres cannot hover stably at close to the minimum temperature.  Fixing the photospheric temperature, we find that the Eddington ratio scales as 
\begin{equation}
{L\over L_E} \propto M_*^{-7/9} M_{\rm BH}^{8/9}
\end{equation}
Thus, growth of the black hole exacerbates the dynamical imbalance, as does partial evaporation of the envelope\footnote{Since the black hole's growth rate is proportional to $M_*$, rapid accretion onto the envelope also would not be able to stabilize the quasistar for very long.} --- the opacity crisis is an unstable situation.  We predict that quasistars entering the forbidden zone must evaporate rather quickly.

There is a complication due to a bump in the opacity from bound-free transitions, which creates a narrow region in the radiative zone where the luminosity is super-Eddington. A similar problem arises in models of Luminous Blue Variables (e.g., Owocki et al.~2004), and remains unresolved.  However, it does not seem likely that this feature alone will lead to catastrophic mass loss in the case of quasistars.
 
\section{Discussion}

The rapid infall of gas in galactic nuclei or pregalactic haloes provides a means for forming seed black holes and rapidly growing them into the intermediate mass regime.  For Pop III quasistars with inflow rates $\sim 0.1-1 M_\odot$ yr$^{-1}$, and simple assumptions about parameters like $\alpha$, the black holes could reach masses $\sim 10^3-10^4 M_\odot$ before the quasistar evaporates.  The quasistar masses could be as large as $10^5-10^6 M_\odot$. Metal-rich quasistars could reach somewhat lower temperatures, but it is not clear whether this implies larger black hole masses, because the run of opacity in the radiative zone would be more complex and dust formation might lead to enhanced mass loss.

Are quasistars detectable? If so, they would be seen at their most massive, shortly before they evaporate. Spectrally, they would resemble 2000--4000 K blackbodies, depending on metallicity, and in the Pop III limit their spectra would be featureless.  Since they radiate at the Eddington limit corresponding to the opacity at the convective--radiative transition --- which is close to that of electron scattering --- their luminosities could reach $10^{43}-10^{44}$ erg s$^{-1}$.  However, their short lifetimes ($\sim 10^5-10^6$ yr) would make them fairly rare.
 
Quasistars can exist only when the envelope mass greatly exceeds the mass of the embedded black hole.  Therefore, they are unlikely to form in the nuclei of galaxies that already possess a supermassive black hole --- in this case, a period of rapid infall (e.g., following a merger) would presumably trigger a quasar outburst instead.  However, low-redshift quasistars might conceivably form in galactic nuclei in which the black hole had been ejected due to three-body interactions, or in which a black hole had never formed.  The earliest plausible sites of quasistar formation would have been pregalactic haloes with virial temperatures exceeding $10^4$ K.  These would have been most common at redshifts $\sim 6-15$.  The spectra of Pop III quasistars at these redshifts would peak at about 10$\mu$m, but they might be marginally detectable by {\it James Webb Space Telescope} on the Wien tail at $3-5\mu$m.  If direct collapse in $10^4$ K haloes is a principal route for forming supermassive black hole seeds, there could be as many as 1--10 per {\it JWST} field, but identifying them would be an extreme challenge.


\begin{theacknowledgments}
The work described here was done in collaboration with Marta Volonteri, Elena Rossi, Martin Rees, and Phil Armitage, and was supported in part by NSF grant AST-0307502, NASA's Beyond Einstein Foundation Science Program grant NNG05GI92G, and a University of Colorado Faculty Fellowship.  The author thanks the Director of the Institute of Astronomy and the Master and Fellows of Trinity College, Cambridge, for their hospitality during a sabbatical visit. 
\end{theacknowledgments}

\bibliographystyle{aipproc}   

\begin{thebibliography}{}

\bibitem{Beg78}
M.~C.~Begelman and M.~J.~Rees, ``The fate of dense stellar systems," MNRAS, 185, 847--860 (1978). 

\bibitem{Beg07}
M.~C.~Begelman, E.~M.~Rossi, and P.~J.~Armitage, ``Quasistars: Accreting black holes inside massive envelopes," MNRAS, submitted (2007).

\bibitem{BVR06}
M.~C.~Begelman, M.~Volonteri, and M.~J.~Rees, ``Formation of supermassive black holes by direct collapse in pre-galactic haloes," MNRAS, 370, 289--298 (2006). 

\bibitem{bla99} 
R.~D.~Blandford and M.~C.~Begelman, ``On the fate of gas accreting at a low rate on to a black hole," MNRAS, 303, L1--L5 (1999).

\bibitem{Bro03} 
V.~Bromm and A.~Loeb, ``Formation of the first supermassive black holes," ApJ, 596, 34--46 (2003).  

\bibitem{DiM05}
T.~Di Matteo, V.~Springel, and L.~Hernquist, ``Energy input from quasars regulates the growth and activity of black holes and their host galaxies," Nature, 433, 604--607 (2005). 

\bibitem{Fan06}
X.~Fan, ``Evolution of high-redshift quasars," New AR, 50, 665--671 (2006).  

\bibitem{Gam01}
C.~F.~Gammie, ``Nonlinear outcome of gravitational instability in cooling, gaseous disks," ApJ, 553, 174--183 (2001). 

\bibitem{Gr98}
A.~Gruzinov, ``The rate of turbulent spherical accretion," unpublished manuscript, astro-ph/9809265 (1998).

\bibitem{Ha61} 
C.~Hayashi, ``Stellar evolution in early phases of gravitational contraction," PASJ, 13, 450--452 (1961). 

\bibitem{HaHo61} 
C.~Hayashi and R.~Hoshi, ``The outer envelope of giant stars with surface convection zone," PASJ, 13, 442--449 (1961). 

\bibitem{Hu75} 
J.~M.~Huntley and W.~C.~Saslaw, ``The distribution of stars in galactic nuclei: Loaded polytropes," ApJ, 199, 328--335 (1975).

\bibitem{Ma05} 
M.~Mayer and W.~J.~Duschl, ``Rosseland and Planck mean opacities for primordial matter," MNRAS, 358, 614--631 (2005).

\bibitem{Na00} 
R.~Narayan, I.~V.~Igumenshchev, and M.~A.~Abramowicz, ``Self-similar accretion flows with convection," ApJ, 539, 798--808 (2000).

\bibitem{owocki04} 
S.~P.~Owocki, K.~G.~Gayley, and N.~J.~Shaviv, A porosity-length formalism for photon-tiring-limited mass loss from stars above the Eddington limit," ApJ, 616, 525--541 (2004).

\bibitem{Qu00} 
E.~Quataert and A.~Gruzinov, ``Convection-dominated accretion flows," ApJ, 539, 809--814 (2000).

\bibitem{Shlos89}
I.~Shlosman, J.~Frank, and M.~C.~Begelman, ``Bars within bars --- A mechanism for fuelling active galactic nuclei," Nature, 338, 45--47 (1989).
 
\bibitem{thorne77} 
K.~S.~Thorne and A.~N.~\.Zytkow, ``Stars with degenerate neutron cores. I. Structure of equilibrium models," ApJ, 212, 832--858 (1977).
 
\bibitem{Wise07}
J.~H.~Wise and T.~Abel, ``Suppression of H$_2$ cooling in the ultraviolet background," ApJ, submitted, astro-ph/0707.2059 (2007). 

\end{thebibliography}

\end{document}